\documentclass[12pt]{article}
\usepackage[utf8]{inputenc}
\usepackage{amsmath}
\usepackage{color}

\begin{document}
\newcommand{\uRbarI}{\stackrel{\overline{\;\;\;\;}\,_I}{{\bf u}_{_R}}}
\newcommand{\uRbari}{\stackrel{\overline{\;\;\;\;}\,i}{{ u}_{_R}}}
\newcommand{\dRbarI}{\stackrel{\overline{\;\;\;\;}\,_I}{{\bf d}_{_R}}}
\newcommand{\dRbari}{\stackrel{\overline{\;\;\;\;}\,i}{{ d}_{_R}}}

\begin{center}
{\LARGE Endowing the Standard Model with a new $r$-symmetry}
\\
\vspace{.3in}
Ngee-Pong Chang\footnote{
email: npccc@sci.ccny.cuny.edu}\\

Physics Department \\
City College of CUNY, New York, NY 10031 \footnote{
Permanent Address }
 \\
and  \\
Institute of Advanced Studies \footnote{
Visiting Professor } \\
Nanyang Technological Univesity, Singapore

\vspace{.2in}
\end{center}

\begin{abstract}

In the Standard Model, there is the single Higgs field, $\phi$, which gives rise to constituent quark and lepton masses.  The Yukawa coupling is a highly complex set of $3 \times 3$ matrices, resulting in many textures of quark and lepton masses.  \\ 

In this talk, we propose to transfer the complexity of the Yukawa coupling matrices to a family of Higgs fields, so that the Yukawa coupling itself becomes a simple interaction. \\

In the context of an Enriched Standard Model, we introduce a new $r$-symmetry in the extended $SU(2)_L \times U(1)_Y \times U(1)_R$ model and show how a particularly simple scenario results in a near degeneracy of masses in the $126-GeV$ region, with a hierarchy of heavier masses suggested by the quark and lepton texture mass matrices.  
\end{abstract}

\begin{flushleft}
\end{flushleft}
\begin{flushleft}
\end{flushleft}

\section{Introduction}

In the Standard Model, the quarks acquire mass through the vacuum expectation value of a single $SU(2)$ Higgs doublet, $\phi^{\alpha}$.  The complexity of the texture of quark mass matrices\cite{texture} is attributed to the Yukawa coupling matrix\footnote{
\samepage
Here $i,j$ range over the values $1, 2, 3$.
We have also suppressed, throughout this paper, the $SU(3)_c$ indices on the quarks.  The Higgs fields are as usual taken to be color neutral.
}

\begin{eqnarray}
  {\cal L}_Y &=& - Y^u_{ij} \uRbari \stackrel{\;\;\;\;\;\;j \,\alpha}{q_{_L}} \epsilon_{\alpha \beta} \;\phi^{\beta} + h.c. \nonumber \\
            && - Y^d_{ij} \dRbari \stackrel{\;\;\;\;\;\;j \,\alpha}{q_{_L}} \phi_{\alpha} + h.c.
\end{eqnarray}
where we have introduced for later convenience the convention
\begin{eqnarray}
		\phi_{\alpha}	&\equiv&  \left( \phi^{\alpha} \right)^{\dagger}
\end{eqnarray}

\noindent Much of the Standard Model phenomenology is dedicated to the determination of the magnitudes and phases of the $CKM$ matrix that can be derived from the $u$, $d$ Yukawa coupling matrices. In particular, there are many unexplained hierarchies in the quark masses, and the Wolfenstein hierarchies in the associated $CKM$ matrices. \\

\noindent In this talk, we propose to transfer the complexity of the Yukawa coupling to the $SU(3)$ family of Higgs fields,
so that the Yukawa interaction is now
\begin{eqnarray}\left.\begin{array}{rcl}
  {\cal L}_Y &=& - h_u \;\uRbari \; \stackrel{\;\;\;\;\;\;j \,\alpha}{q_{_L}} \epsilon_{\alpha \beta} \;\phi^{\beta}_{_{ij}} + h.c.  \\
            && - h_d \;\dRbari \; \stackrel{\;\;\;\;\;\;j \,\alpha}{q_{_L}}  \;\widehat{\phi}\,{}_{ij \,\alpha} + h.c.
            \end{array} \right\} \label{Yukawa u-d int}
\end{eqnarray}
Note that in so doing we have also added the distinction between the Higgs fields, $\widehat{\phi}\,{}_{ij \,\alpha}$, coupled to the down-quark family versus the original  $\phi^{\alpha}_{ij}$ associated with the up-quark family.  From a certain point of view, it looks counter-intuitive to double down on the number of Higgs fields when phenomenologically there was such a great difficulty in even finding one Higgs.  The pay-off comes when you realize there is a greater symmetry that results from so doing.

\section{r-Symmetry}

To make the model more attractive, we make the simple requirement 
\begin{eqnarray}
		h_u	&=& h_d
\end{eqnarray} 
so that
\begin{eqnarray}
{\cal L}_Y &=& -h_q \left( \stackrel{\overline{\;\;\;\,}\,_i}{ u_{_R}} \;q_{L}^{j \,\alpha} \;\phi^{\beta}_{_{ij}} 
					\;\epsilon_{\alpha \beta}   \;+\; \stackrel{\overline{\;\;\;\,}\,_i}{ d_{_R}} \;q_{L}^{j \,\alpha} 
					\; \widehat{\phi}\,{}_{_{ij}\alpha} \;\;\right) \;\;\;\;+ \; h.c. 
\end{eqnarray}

\noindent This requirement may look odd, as everyone knows that the down quark in each family is much lighter than the up quark.  But, for this Enriched Standard Model, with the two families of Higgs fields, $\phi^{\alpha}$, and $\widehat{\phi}\,{}_{\alpha}$, the difference in the physical up and down quark masses may be attributed to the difference in vacuum expectation values of the corresponding Higgs fields. \\

\noindent To implement this requirement we extend the enriched Standard Model to the gauge group $SU(2)_L \times U(1)_Y \times U(1)_R$ and impose a new $r$-symmetry on the full Lagrangian.   \\

\begin{center}
\begin{tabular}{|c|c|c|c|| c| c|c| c |}
		\hline
			& $Y_R$	&	$Y'$ & $(I_{3})_{L}$   & & $Y_R$	&	$Y'$ & $(I_{3})_{L}$  \\  \hline
		$\phi\,{}^{+}_{ij}$	&	$+1/2$	&	& $+ 1/2$ 	 & $u^{j}_{L}$ &  & $+1/6$ & $+1/2$ \\
		$\phi\,{}^{o}_{ij}$	&	$+ 1/2$ & & $-1/2$ 		 & $d^{j}_{L}$ &  & $+1/6$ & $-1/2$\\
		$\widehat{\phi}\,{}^{+ \,ij}$	& $+ 1/2$ 	 &  & $+ 1/2$ & $u_{_{Ri}}$ & $+1/2$  & $+1/6$ &  \\
		$\widehat{\phi}\,{}^{o \,ij}$  &	$+ 1/2$		 &  & $- 1/2$ & $d_{_{Ri}}$ & $-1/2$  & $+1/6$ & \\
		\hline
\end{tabular}
\end{center}

\noindent The covariant derivatives for the quark \& Higgs fields now read as\footnote{ 
		Here $a,b = 1,2$ refer to the flavor of the right-handed quark families $q^{a}_{_R \,i}$, while $\alpha = 1,2$ refer as usual to 
		the flavor of the left-handed $q^{j}_{L}$ quark families, and  $i,j=1,2,3$ refer to the families. 
}
\begin{eqnarray}
	D_{\mu} \;q^{j \,\alpha}_L	&=& \partial_{\mu} \;q^{j \,\alpha}_{L} + i \;\frac{g}{2} \;\left( \vec{\tau} \cdot \vec{W} \right)^{\alpha}_{\beta} \;q^{j \,\beta}_{L} + i \;\frac{g'}{6} \;B^{'}_{\mu} \;q^{j \,\alpha}_{L}  \\
	D_{\mu} \;q^{a}_{R\,i}	&=& \partial_{\mu} \;q^{a}_{R\,i} + i \;\frac{g_R}{2} \;\Bigl( \tau_3 \;B_{R \,\mu} \Bigr)^{a}_{b} \;q^{b}_{R\,i} + i \;\frac{g'}{6} \;B^{'}_{\mu} \;q^{a}_{R\,i} \\
	D_{\mu} \,\phi^{\alpha}_{ij} &=& \partial_{\mu}\,\phi^{\alpha}_{ij} \,+\, i \,\frac{g}{2} \,\left( \vec{\tau} \cdot \vec{W} \right)^{\alpha}_{\beta} \,\phi^{\beta}_{ij}  \,+\, i \,\frac{g_R}{2} \;B_{R\mu} \,\phi^{\alpha}_{ij}  \\
	D_{\mu} \,\widehat{\phi}\,{}_{\alpha \,ij} &=& \partial_{\mu}\,\widehat{\phi}\,{}_{\alpha \,ij} \,-\, i \,\frac{g}{2} \,\left( \vec{\tau} \cdot \vec{W} \right)_{\alpha}^{\beta} \,\widehat{\phi}\,{}_{\beta \,ij}  \,-\, i \,\frac{g_R}{2} \;B_{R\mu} \,\widehat{\phi}\,{}_{\alpha \,ij} 
\end{eqnarray}

\noindent The full Lagrangian is invariant under the $r$-symmetry

\begin{eqnarray}
\left. 
\begin{array}{rclcrcl}
	u_{_{Ri}}		& \rightarrow &  \;\;\;\;\;\; d_{_{Ri}}  & & 	d_{_{Ri}}		& \rightarrow & - u_{_{Ri}} \\
	\phi\,{}^{\alpha}_{ij} & \rightarrow & - \epsilon^{\alpha \beta} \;\widehat{\phi}\,{}_{\beta \,ij} & &
	\widehat{\phi}\,{}_{\alpha \,ij} & \rightarrow & - \epsilon_{\alpha \beta} \;\phi\,{}^{\beta}_{ij}
\end{array}
	\right\} &&
\end{eqnarray}
and for the gauge fields
\begin{eqnarray}
\left. \begin{array}{ccl}
	B_R \,{}_{\mu} &\rightarrow& - \; B_R \,{}_{\mu}  \\
	\vec{W}_{\mu}	&\rightarrow& + \;\vec{W}_{\mu} \\
	B^{'}_{\mu}	&\rightarrow&  + \;B^{'}_{\mu}	
	\end{array} \right\} &&
\end{eqnarray}

\noindent This extension of the Standard Model parallels the $SU(2)_{L} \times SU(2)_R \times U(1)_Y$ of Mohapatra \& Senjanovic\cite{Mohapatra}.  Instead of the full set of $SU(2)_R$ gauge bosons, however, we have only the neutral $B_R$ gauge boson.
To have the correct neutrino phenomenology, we introduce a set of very heavy Higgs fields, $ \Delta^{R \,,ij}_{ab}$.

\section{Include Leptons}

\noindent To include leptons, we introduce the compact notation for the Higgs fields
\begin{eqnarray}
\phi^{\beta}_{a \,,ij} &=& \left(  
		\begin{array}{rcc}
	\widehat{\phi}\,{}_{(o) \,ij} && \phi^{(+)}_{ij}   \\
	- \widehat{\phi}\,{}_{(-) \,ij} &&  \phi^{(o)}_{ij} 
		\end{array}  \right)^{\beta}_{a} 
\end{eqnarray}
using the convention that
\begin{eqnarray}
\left.  	\begin{array}{lcl}
	\phi^{ij}_{\alpha}  &\equiv&  \left( \phi^{\alpha}_{ij} \right)^{\dagger}  \\ 
	\widehat{\phi}\,{}^{\alpha \,ij}  &\equiv&  \left( \widehat{\phi}_{\alpha \,ij} \right)^{\dagger}  
					\end{array} \right\}
\end{eqnarray}

\noindent Here in addition to the Dirac mass terms for the leptons, we introduce the Majorana mass terms for the leptons.  The complete fermion Yukawa Lagrangian now reads
\begin{eqnarray}
\left.  	\begin{array}{rcl}
	{\cal L}_{Y} &=& \;-\; h_{q} \; \stackrel{\overline{\;\;\;\;\;}\;\;_i}{{ q}_{_{R \,a}}} \;q^{j \alpha}_{L} \;\epsilon_{\alpha \beta} \;\phi^{\beta}_{b \,ij} \;\epsilon^{ab} \;+\; h.c.  \\
			& & \;-\; h_{\ell} \; \stackrel{\overline{\;\;\;\;}\;_i}{{ \ell}_{R \,a}} \;\ell^{j \alpha}_{L} \;\epsilon_{\alpha \beta} \;\phi^{\beta}_{b \,ij} \;\epsilon^{ab} \;+\; h.c.  \\
			& & \;+\; \frac{1}{2}  \;\widetilde{\ell}^{\,a}_{R \,i} \;{\cal C} \;\ell^{\,b}_{R \,j} \;\Delta^{R \,,ij}_{ab} \;\;+\; h.c.  
					\end{array} \right\}
\end{eqnarray}
Under $r$-symmetry, the fermion fields transform as ( $\left( i \sigma_2 \right)^a_b \;=\; \epsilon_{ab}$ )
\begin{eqnarray}
\left.  	\begin{array}{rcl}
		q^{a}_{_{R \,i}}		& \rightarrow &  \left( i \sigma_2 \right)^{a}_{b} \;q^{b}_{_{R \,i}} \\
		q^{j \,\alpha}_{L}	& \rightarrow &  q^{j \,\alpha}_{L} \\
		\ell^{a}_{R \, i} &\rightarrow& \left( i \sigma_2 \right)^{a}_{b} \;\ell^{b}_{R \,i} \\
		\ell^{j \,\alpha}_{L} & \rightarrow & \ell^{j \,\alpha}_{L}
					\end{array} \right\}
\end{eqnarray}
while the Higgs fields transform as
\begin{eqnarray}
\left.  	\begin{array}{rcl}
		\phi^{\beta}_{b \,ij}	&\rightarrow&  \left( - i\sigma_2 \right)^{c}_{b} \;\phi^{\beta}_{c \,ij}  \\
		\Delta^{R \,ij}_{ab} &\rightarrow& \left( i \sigma_2 \right)^{c}_{a} \;\Delta^{R \,ij}_{cd} 
						\;\left( i \sigma_2 \right)^{d}_{b} 
					\end{array} \right\}
\end{eqnarray}
or, specifically for the new $\Delta^{R \,,ij}_{ab}$ fields under $r$-symmetry
\begin{eqnarray}
\left.  	\begin{array}{rcl}
		\Delta^{R \,ij}_{11}  & \longleftrightarrow & \;\;\Delta^{R \,ij}_{22}  \\
		\Delta^{R \,ij}_{12}  & \longleftrightarrow & - \Delta^{R \,ij}_{21} 
					\end{array} \right\}
\end{eqnarray}

\noindent The covariant derivatives of the Higgs fields now read in totality
\begin{eqnarray}
	\left.  \begin{array}{lcl}
	D_{\mu} \phi^{\beta}_{ij \,a} &=&  \partial_{\mu} \phi^{\beta}_{ij \,a} + i \displaystyle \frac{g}{2} \left( \vec{\tau} \cdot \vec{W}_{\mu} \right)^{\beta}_{\alpha} \phi^{\alpha}_{ij \,a}  - i \frac{g_{_R}}{2} \left( \sigma_3 \right)^{a'}_{a} B_{_R \,\mu} \phi^{\alpha}_{ij \,a'}      \\
	D_{\mu} \Delta^{ij}_{R \,ab } &=&  \partial_{\mu} \Delta^{ij}_{R \,ab } - i \displaystyle \frac{g_{_R}}{2} \left( \sigma_3 \right)^{a'}_{a} B_{_R \,\mu}\Delta^{ij}_{R \,a'b }  \\
		&&  - i \displaystyle \frac{g_{_R}}{2} \left( \sigma_3 \right)^{b'}_{b} B_{_R \,\mu} \Delta^{ij}_{R \,ab' } 
				+ i g' B_{\mu} \Delta^{ij}_{R \,ab }    
				\end{array}  \right\} && \label{covar deriv}
\end{eqnarray}

\noindent Following Mohapatra \& Senjanovic\cite{Mohapatra}, we implement the symmetry breaking pattern such that the vacuum expectation values of the different Higgs fields 
\begin{eqnarray}
	< \Delta^{ij}_{R \,ab} > &=& \left(  
			\begin{array}{ccc}
			\upsilon_R^{ij} && 0   \\
			0 &&  0
			\end{array}  \right)_{ab} \\
	 < \phi^{\beta}_{b \,ij} > &=& \left(
			\begin{array}{ccc}
			\widehat{\upsilon}\,{}_{ij} & 0 \\
			0 & \upsilon_{ij} 
			\end{array} \right)^{\beta}_{b}
\end{eqnarray}
possess the hierarchy
\begin{eqnarray}
\upsilon_{R}  \;>> \; \upsilon \;>>\; \widehat{\upsilon} 
\end{eqnarray}
where we have used the simplified notation
\begin{eqnarray}
\left.  \begin{array}{lcl}
		\upsilon^2	&=& \upsilon^{ij} \;\upsilon_{ij}  \\
		\widehat{\upsilon}\,{}^2	&=& \widehat{\upsilon}\,{}_{ij} \;\widehat{\upsilon}\,{}^{ij}  \\
		\upsilon^2_{R}	&=& \upsilon^{ij}_{_R} \;\upsilon_{ij \,_R}
				\end{array} \right\}  \label{vev}
\end{eqnarray}

\section{Gauge Bosons Acquire Mass}
\noindent From the covariant derivatives in eq.(\ref{covar deriv}), we arrive at the physical gauge bosons ( $ \upsilon^2_{t} \;\equiv\; \upsilon^2 + \widehat{\upsilon}\,{}^2$ )

\begin{eqnarray}
	\left.   \begin{array}{lcl}
	Z_{\mu} &=& \cos{\theta_W} \;W^{3}_{\mu}  \;-\; \sin{\theta_W} \; \displaystyle\frac{\left( g' \,B_{R \mu} + g_{_R} \,B_{\mu} \right)  }{ \sqrt{g_{_R}^2 + g'\,{}^2} } \;+\; O \Bigl( \frac{\upsilon_{t}^2}{\upsilon_R^2} \Bigr)  \\
	A_{\mu} &=& \sin{\theta_W} \;W^{3}_{\mu} \;+\; \cos{\theta_W} \; \displaystyle \frac{\left( g' \,B_{R \mu} + g_{_R} \,B_{\mu} \right) } {\sqrt{g_{_R}^2+ g'\,{}^2}}  \\
	Z_{R \mu} &=& \displaystyle \frac{g_{_R} \,B_{R \mu} - g' \,B_{\mu}}{\sqrt{g_{_R}^2 + g'\,{}^2}}  \;+\; O \Bigl( \frac{\upsilon_{t}^2}{\upsilon_R^2} \Bigr)  
							\end{array} \right\} && \label{gauge boson}
\end{eqnarray}
with the masses given by
\begin{eqnarray}
	m_W^2 &=& \frac{1}{2} \,g^2 \upsilon_t^2    \\
	m_Z^2 &=& \displaystyle \frac{m_W^2}{\cos^2 \theta_W }   \\
		&=& \frac{\upsilon^2_t}{2} \; \left( g^2 + g\,{}^{\prime \,2}_s \right)  \label{m_Z std} \\
	m_{Z_R}^2 &=& 2 \left( \upsilon_R^2  \right) \;\left( g_{_R}^2 + g\,{}^{\prime \,2}_s \right) \;+\; O(\upsilon^2) 
\end{eqnarray}
where
\begin{eqnarray}
	g\,{}^{\prime}_s	&=&  \displaystyle \frac{ g_{_R} \,g\,{}^{\prime}}{ \sqrt{ g^2_{_R} + g\,{}^{\prime \,2}_s   } } 
\end{eqnarray}
or
\begin{eqnarray}
	\displaystyle \frac{1}{g\,{}^{\prime \,2}_s }  &=& \frac{1}{g^2_{_R} } + \frac{1}{ g\,{}^{\prime \,2} }  \label{g's relation}
\end{eqnarray}
From eq.(\ref{m_Z std}) we see that the $g'_{s}$ is actually the $U(1)_Y$ coupling of the Standard Model, and eq.(\ref{g's relation}) gives its relationship with the couplings of the extended $SU(2)_L \times U(1)_Y \times U(1)_R$ model.  \\

\noindent These relationships are a manifestation of the decoupling theorem of Georgi-Weinberg\cite{Georgi-Weinberg}.  In eq.(\ref{gauge boson}) we see how in the limit of $ \upsilon_{t} / \upsilon_R \rightarrow 0$, 
\begin{eqnarray}
		\displaystyle \frac{ g' B_{R \mu} + g_{_R} B_{\mu} }{ \sqrt{ g^2_{_R} + g'^2 }} &\rightarrow&  B_{Y\mu}
\end{eqnarray}
where $B_{Y \mu}$ is the $U(1)_Y$ gauge field of the usual $SU(2)_L \times U(1)_Y$ group.

\section{A Simple Higgs Potential}

As noted already by Mohapatra and Senjanovic\cite{Mohapatra}, neutrino phenomenology requires that the Higgs fields $\Delta^{ij}_{R \,ab}$ be associated with a mass scale that is much higher than the Higgs $\phi^{\beta}_{a \,ij}$.  For our purposes, the Georgi-Weinberg decoupling theorem\cite{Georgi-Weinberg} enables us to focus on the low energy phenomenology associated with the $\phi^{\beta}_{a \,ij}$.  \\

\noindent Rather than work with a most general for the Higgs potential, We turn to a particularly simple form for the Higgs potential.  \\

\noindent We consider the potential
\begin{eqnarray}
		V &=& V_{\phi} + V_{\Delta}    \label{Higgs Full Pot}
\end{eqnarray}
where $V_{\phi}$ involves the lighter $\phi^{\alpha}_{ij}$ and $\widehat{\phi}\,^{\alpha \,k\ell}$ fields, while $V_{\Delta}$ involves the heavy $\Delta^{ij}_{R \,ab}$ fields.  
For the general non-degenerate case, we introduce the three coupling constants, $\lambda_1, \lambda_2, \lambda_3$ in the maximally symmetric potential
\begin{eqnarray}
		V_{\phi}	&=& + \displaystyle \frac{\lambda_1}{2} \left( \epsilon_{\alpha \beta} \;\phi\,{}^{\alpha}_{a, \,ij} \;\phi\,{}^{\beta}_{b, \,k\ell}  \right) 
		\left( \epsilon^{\alpha' \beta'} \;\phi\,{}_{\alpha'}^{a, \,ij} \;\phi\,{}_{\beta'}^{b, \,k\ell}  \right)  \nonumber \\
							& & + \displaystyle \frac{\lambda_2}{2} \left( \phi\,{}^{\alpha}_{a, \,ij} \;\phi\,{}^{b, \,k\ell}_{\alpha} \right) 
		\left( \phi\,{}^{\beta}_{b, \,k\ell} \;\phi\,{}^{a, \,ij}_{\beta} \right) - \displaystyle \frac{\lambda_3}{4} \left( \phi\,{}^{\alpha}_{a, \,ij} \;\phi\,{}^{a, \,ij}_{\alpha} \right)^2   
\end{eqnarray}
The symmetry is broken through the vacuum expectation values of the $\phi\,{}^{\beta}_{a, \,ij}$ fields, as given in eq.(\ref{vev}) above.  \\

\noindent The Higgs potential $V_{\phi}$ around the new vacuum takes the form
\begin{eqnarray}
		V_{\phi}	&=& + \displaystyle \frac{\lambda_1}{2} \left( \epsilon_{\alpha \beta} 
										\left[ \;\phi\,{}^{\alpha}_{a, \,ij} \;\phi\,{}^{\beta}_{b, \,k\ell} 
												- \upsilon\,{}^{\alpha}_{a, \,ij} \;\upsilon\,{}^{\beta}_{b, \,k\ell} \right] \right) \times \nonumber \\
							& &   \;\;\;\;\;\;\; \left( \epsilon^{\alpha' \beta'} \left[ \;\phi\,{}_{\alpha'}^{a, \,ij} \;\phi\,{}_{\beta'}^{b, \,k\ell} 
												- \upsilon\,{}_{\alpha}^{a, \,ij} \;\upsilon\,{}_{\beta}^{b, \,k\ell} \right]  \right)  \nonumber \\
							& & + \displaystyle \frac{\lambda_2}{2} 
										\left( \left[ \phi\,{}^{\alpha}_{a, \,ij} \;\phi\,{}^{b, \,k\ell}_{\alpha}  
												- \upsilon\,{}^{\alpha}_{a, \,ij} \;\upsilon\,{}^{b, \,k\ell} \right] \right) \times  \nonumber \\
							& &   \;\;\;\;\;\;\; \left( \left[ \phi\,{}^{\beta}_{b, \,k\ell} \;\phi\,{}^{a, \,ij}_{\beta} 
												- \upsilon\,{}^{\beta}_{b, \,k\ell} \;\upsilon^{a, \,ij}_{\beta} \right]  \right) \nonumber \\
							& & - \displaystyle \frac{\lambda_3}{4} \Bigl( \phi\,{}^{\alpha}_{a, \,ij} \;\phi\,{}^{a, \,ij}_{\alpha} 
												- \upsilon^2 - \widehat{\upsilon}\,{}^2 \Bigr)^2   \label{Higgs Pot}
\end{eqnarray}

\noindent Likewise, the Higgs potential involving the heavy fields takes the form
\begin{eqnarray}
\left.		\begin{array}{lcl}
			V_{\Delta} &=& +  \;\lambda_4 \;\left( \Delta^{ij}_{R \,ab} \,\Delta^{R \,ab}_{ij} \;-\; \upsilon^2_{R}  \right)^2  \\
					& & +  \;\lambda_5 \;\left( \Delta^{ij}_{R \,ab} \,\Delta^{k\ell}_{R \,cd} \;\epsilon^{ac} 
									\epsilon^{bd} \right)  
									\,\left( \Delta_{ij}^{R \,a'b'} \,\Delta_{k\ell}^{R \,c'd'} \;\epsilon_{a'c'} \epsilon_{b'd'} \right)\\
					& & +  \;\lambda_6 \;\upsilon^2_{R} \left[  \Delta^{ij}_{R \,ab} \left( \sigma_1 \right)^{a}_{a'}  \Delta^{R \,a'b'}_{ij} 
					\left( \sigma_1 \right)^{b}_{b'}  \right. \\
					& &\left. \;\;\;\;\;\;\;\;\;\;\; -  \Delta^{ij}_{R \,ab} \left( i \sigma_2 \right)^{a}_{a'}  \Delta^{R \,a'b'}_{ij} 
						\left( i \sigma_2 \right)^{b}_{b'}   \right]
				\end{array} \right\}
				\label{Higgs Pot_'}
\end{eqnarray}

\noindent  By construction, this Higgs potential is stable about the broken vacuum with $\upsilon_{ij}$ and $\widehat{\upsilon}\,{}_{k\ell}$.

\section{Leading Hierarchy}

\noindent The Higgs potential in eq.(\ref{Higgs Pot}) gives rise to a rich mass spectrum.  It involves the full complexity of the texture of the vacuum expectation values, and poses a daunting task for the timid explorer. Fortunately, the observed hierarchy in the quark masses, with $m_t >> m_b >> m_c >> m_s >> m_u$, enables us to explore the spectrum in the leading hierarchy\footnote{\samepage
On a technical note, this hierarchy in the vacuum expectation values is manifest only in the fermion mass diagonal basis. As the Higgs potential is manifestly invariant in the family indices $i,j$, we will carry through the subsequent discussion of the Higgs mass hierarchy in this basis. 
}
, where we simply set $\upsilon_{33} = \upsilon, \widehat{\upsilon}\,{}_{33} = \widehat{\upsilon}$, and let all the other vacuum expectation values be zero.  \\

\noindent We express the resulting spectrum in terms of the hermitian fields, $h$ and $z$, defined by
\begin{eqnarray}
		\phi^{o}_{33}	&=&  \displaystyle \frac{h_{3} - i \;z_3}{\sqrt{2}}   \\
		\widehat{\phi}\,{}^{o \,33}	&=& \displaystyle \frac{\widehat{h}\,{}_{3} - i \;\widehat{z}\,{}_3}{\sqrt{2}}  
\end{eqnarray}
and for $(ij) \neq (33)$
\begin{eqnarray}
		\phi^{o}_{ij}	&=&  \displaystyle \frac{h_{ij} - i \;z_{ij}}{\sqrt{2}}   \\
		\widehat{\phi}\,{}^{o \,ij}	&=& \displaystyle \frac{\widehat{h}\,{}_{ij} - i \;\widehat{z}\,{}_{ij}}{\sqrt{2}}  
\end{eqnarray}

\section{Goldstone Bosons}
The total Higgs potential, eq.(\ref{Higgs Full Pot}), leaves as massless the usual Goldstone bosons of the $SU(2)_L \times U(1)_Y \times U(1)_R$ gauge theory
\begin{eqnarray}
		\left. \begin{array}{lcl}
					h^{+}_A	&\equiv&	\displaystyle \frac{\upsilon \,\phi^{(+)}_{33} + \widehat{\upsilon} \,\widehat{\phi}\,{}^{(+) \,33} }{ \sqrt{\upsilon^2 + \widehat{\upsilon}\,{}^2} } \vspace{.1in} \\
					z_A			&\equiv&	\displaystyle \frac{i}{\sqrt{2}} \Bigl( \displaystyle \frac{\upsilon \,\phi^{o}_{33} + \widehat{\upsilon} \,\widehat{\phi}\,{}^{o \,33} } { \sqrt{\upsilon^2 + \widehat{\upsilon}\,{}^2} } \;-\; h.c. \Bigr) \vspace{.1in} \\
					z_{R}  &\equiv&	\displaystyle \frac{i}{\sqrt{2}} \Bigl( \Delta_{33}^{R \,11} \;-\; h.c. \Bigr)
						\end{array} \right\}
\end{eqnarray}
In the 't Hooft gauge, they acquire mass through the gauge-fixing terms for the massive gauge bosons, $W^{\pm}_{\mu}$, $Z_{\mu}$ and $Z_{R \mu}$, so that
\begin{eqnarray}
	{\cal L}_{mass}	&=& \;\;-\displaystyle \frac{g^2_R}{\cos^2{\theta_R}} 
			\;\Bigl( z_{R \,ij} \;\upsilon_{R}^{ij} + \frac{\cos^2{\theta_R}}{2} \left( z_{ij} \;\upsilon^{ij} 
				+ \widehat{z}\,{}^{ij} \;\widehat{\upsilon}\,{}_{ij} \right)  \Bigr)^2 \nonumber \\
		& & - \frac{g^2}{4 \cos^2{\theta_W }} \;\Bigl( \left( z_{ij} \;\upsilon^{ij} 
				+ \widehat{z}\,{}^{ij} \;\widehat{\upsilon}\,{}_{ij} \right) 
				- \frac{\upsilon^2_{t}}{2 \upsilon^2_{R}} \cos^2{\theta_R} \;\left[ z_{R \,ij} \;\upsilon_R^{ij} \right]  \Bigr)^2 \nonumber  \\
		& & - \left|  \;\frac{g}{\sqrt{2}} \left( \;\upsilon^{ij} \;\phi^{(+)}_{ij} 
				\;+\; \widehat{\upsilon}\,{}_{ij} \;\widehat{\phi}\,{}^{(+)\,ij} \right)  \right|^2 \nonumber
\end{eqnarray}
and the Goldstone Boson states that acquire mass are the linear combinations
\begin{eqnarray}
		z_A	&=& {\cal N} \Bigl( \displaystyle \frac{ \left( \upsilon \,z_3 
						+ \widehat{\upsilon} \,\widehat{z}\,{}_3 \right) }{\upsilon_t}
						- \frac{\upsilon_t \cos^2{\theta_R}}{2 \upsilon_R} \;z_R  \Bigr)  \\
		z^R_A &=& {\cal N} \Bigl( z_R + \frac{\cos^2{\theta_R}}{2 \upsilon_R} \,\left( \upsilon \,z_3 
						+ \widehat{\upsilon} \, \widehat{z}\,{}_3 \right)  \Bigr) 
\end{eqnarray}
with the normalization
\begin{eqnarray}
		{\cal N}	&=& \displaystyle \frac{1} {\sqrt{ 1 + \; \upsilon^2_t \cos^4{\theta_R} /( 4 \upsilon^2_R )   } }
\end{eqnarray}
where
\begin{eqnarray}
	\cos{\theta_R} &=&  \frac{g_{_R}}{ \sqrt{g_{_R}^2 + g'\,{}^2} }   
\end{eqnarray}
and we have the masses
\begin{eqnarray}
\left.   \begin{array}{lcl}
	m ( h^{+}_A )	&=&  M_W	\\
	m ( z_{A} )		&=&	 M_Z	\\
	m ( z^{R}_{A} ) 	&=& M_{Z_R}
				\end{array}
					\right\}
\end{eqnarray}

\section{Leading Hierarchy Mass Spectrum}

The scale $\upsilon_{R}$ is super heavy compared with the lighter Higgs scale $\upsilon, \widehat{\upsilon}$.  In what follows, we  focus on the physical mass spectrum that arise in the Higgs potential, $V_{\phi}$, in eq.(\ref{Higgs Pot}) .

\begin{center}
\begin{tabular}{|c|| l|}
		\hline 
			$M^2$	&	 $\;\;\;\;\;\;\;\;\;\;\; Higgs$  \\  
			\hline
		$( \lambda_1 + \lambda_2 - \lambda_3) \upsilon^2_t $	&	  
								$\;\;\;\;\; h_A = (\upsilon \,h_3 + \widehat{\upsilon} \,\widehat{h}\,{}_3 )/\upsilon_t $ \\
		$\lambda_2 \,\upsilon^2_t $ & $\left\{ \begin{array}{lcl}
																		h_B &=& ( \widehat{\upsilon} \,h_3 - \upsilon \,\widehat{h}\,{}_3 )/\upsilon_t \\
																		z_B &=& ( \widehat{\upsilon} \,z_3 - \upsilon \,\widehat{z}\,{}_3 )/\upsilon_t \\
																		\phi^{+}_B &=& ( \widehat{\upsilon} \,\phi^{+}_{33} 
																									- \upsilon \,\widehat{\phi}\,{}^{+ \,33} )/\upsilon_t 
																					\end{array} \right. $ \\
		$ \lambda_1 \widehat{\upsilon}\,{}^2 + \lambda_2 \,\upsilon^2$ & $\;\;\;\;\; h_{ij} \;, \;z_{ij}\;, \; \widehat{\phi}\,{}^{+ \,ij}  \;\;\;{\rm for} \;ij \neq 33 $ \\
		$ \lambda_1 \upsilon^2 + \lambda_2 \,\widehat{\upsilon}\,{}^2$ & $\;\;\;\;\; \widehat{h}\,{}_{ij} \;, \;\widehat{z}\,{}_{ij}\;, \; \phi^{+ \,ij}  \;\;\;{\rm for} \;ij \neq 33 $ \\
		\hline
\end{tabular}
\end{center}

\section{Toy Model}

While a full analysis of the phenomenology of an Enriched Standard Model should allow for the possibility of a range of values for the coupling constants $\lambda_1, \lambda_2, \lambda_3$, we will consider in this talk the rather intriguing possibility that the three coupling constants are equal. 
\begin{eqnarray}
\begin{array}{ccccccc}
	\lambda_1 &=& \lambda_2 &=& \lambda_3 &=& \lambda
\end{array}			\label{degen lambda}
\end{eqnarray}

\noindent With this choice of a common coupling constant, the resulting spectrum of the Higgs family of physical states are \underline{\bf all degenerate} in the leading hierarchy !  \\

\noindent In a full calculation of the mass spectrum, the texture of the $\upsilon_{ij}$ and $\widehat{\upsilon}\,{}_{ij}$ matrices will lead to predictable splittings of the degeneracy.  \\

\section{Phenomenological Implications}

\noindent There has been a lot of excitement in the world of particle physics ever since the announcement of the discovery of Higgs boson on July 4, 2012.  Now that we contemplate an Enriched Standard Model, a natural question that arises would be which of the family of Higgs is the one reported at $125.9 \;\pm\; 0.4 \;GeV$ ?  The answer becomes clear when we take into account the production and decay channels involved. \\

\noindent For at high energies, the leading production processes predominantly involve top quark loops.  Therefore, the production of the family Higgs, $h_{ij}$, $z_{ij}, \;\phi^{+}_{ij}, \;\widehat{\phi}\,{}^{+}_{ij}$, with $(ij) \neq (33)$, are suppressed. \\

\noindent Among the family of Higgs, $h_A$ plays a special role.  It behaves like the single Higgs field of the Standard Model, with the same trilinear coupling to the gauge bosons as the standard Higgs.  It is produced via associative production with $W$ or Vector Boson fusion.  Its production cross-section is identical to that of the Standard Model.  \\

\noindent In contrast, the orthogonal Higgs bosons, $h_B$ and $z_B$, do not have trilinear couplings to the gauge bosons.  They are produced through coupling to the top and bottom quarks, see eq.(\ref{Higgs fermion coupling}) below.  While their mass is degenerate with the $h_A$ boson, they have different decay widths into $b \bar{b}$ channel.

\section{Higgs couplings to fermions}
For completeness, we list here the couplings of the neutral Higgs mesons to the fermions
\begin{eqnarray}
\left. 	\begin{array}{lcl}
	{\cal L}_{Y} &=& - \;\displaystyle \frac{h_t}{\sqrt{2}} \;\bar{t} \;t \;\frac{\left( \upsilon \, h_A + \widehat{\upsilon}\, h_B \right)}{\upsilon_t}   \;+\; i \;\frac{h_t}{\sqrt{2}} \;\bar{t} \;\gamma_5 \;t \; \frac{ \left( \upsilon \;z_A + \widehat{\upsilon}\, z_B \right) }{\upsilon_{t}}  \\
		&&  -  \;\displaystyle \frac{h_t}{\sqrt{2}} \;\bar{b} \;b \;\frac{\left( \widehat{\upsilon}\, h_A - \upsilon \, h_B \right)}{\upsilon_t}   \;-\; i \;\frac{h_t}{\sqrt{2}} \;\bar{b} \;\gamma_5 \;b \; \frac{ \left( \widehat{\upsilon}\, \;z_A - \upsilon \, z_B \right) }{\upsilon_{t}} 
				\end{array} \right\} 	\label{Higgs fermion coupling}
\end{eqnarray}
where $z_A$ are the Goldstone bosons in the 't Hooft gauge with mass $M_Z$.
Note that here $h_A$ and $h_B$ are scalar fields, while $z_B$ is a pseudoscalar field.

\section{Parameters}

We focus only on the $j=3$ family, and use as input the observed values
\begin{eqnarray}
\left.  
\begin{array}{lcll}
	m_W	&=& 80.39 &GeV \\
	m_Z	&=& 91.19 &GeV \\
	\upsilon_{std} &=& 173.95 &GeV   \\
	m_t	&=& 173.07 &GeV \\
	m_b	&=& 4.19 &GeV  \\
 	m_H 	&=& 125.9 \pm 0.4 & GeV
\end{array}
	\right\} && 
\end{eqnarray}
From the observed quark and charged vector boson masses
\begin{eqnarray}
\left. \begin{array}{lcl}
	m_t	&=&  h_q \;\upsilon  \\
	m_b	&=&  h_q  \;\widehat{\upsilon} 
			\end{array} \right\} &&
\end{eqnarray}
we find the parameters
\begin{eqnarray}
\left. \begin{array}{lcl}
	\upsilon	&=& 173.9 \;GeV \\
	\widehat{\upsilon}	&=& 4.2 \;GeV \\
	h_t				&=& 0.995 \\
	\lambda		&=& 0.524 
				\end{array} \right\} && 
\end{eqnarray}
with the Higgs masses set at $125.9 \;\pm \;0.4 \;GeV$. \\

\noindent  Based on these parameters, we arrive at the zeroth-order decay widths
\begin{eqnarray}
\left.  \begin{array}{lcl}
	\Gamma_{h_A \rightarrow b \bar{b}} &=&  0.004 \;GeV  \\
	\Gamma_{h_B \rightarrow b \bar{b}} &=&  7.389 \;GeV  \\
	\Gamma_{z_B \rightarrow b \bar{b}} &=&  7.422 \;GeV  
				\end{array} \right\}
\end{eqnarray}
Note that these widths are in the leading hierarchy approximation, and do not include the degeneracy lifting effects arising from the texture of the quark matrices.  But they do hint at a rich structure in the decay width into $b \bar{b}$ mode.

\section{Conclusion}

The discovery of Higgs at LHC is truly a tribute to the perseverance and dedication of decades of experimental effort at confirming the Standard Model.  Having found this $125.9 \pm 0.40 \;GeV$ peak, it is tempting to immediately identify it with the lone Higgs field, $\phi$, in the Standard Model (SM). \\

\noindent Out of curiosity, we can ask if there are degenerate masses hidden in the peak, and, if so, how many.  Rather than as an idle question, what I have tried to describe in this talk, is a framework where interesting scenarios do arise.  I have proposed endowing the SM with a family of Higgs fields $\phi_{ij}$, $\widehat{\phi}\,{}^{ij}$, so that the Yukawa coupling is simplified.  To bring some order to the resulting Higgs potential, we have also introduced a new $r$-symmetry.  \\

\noindent The resulting hierarchy of masses among the Higgs family is remarkable.  This enriched Standard Model gives rise to a scenario where the $125.9 \pm 0.40 \;GeV$ peak\cite{LHC} may be resolved into a triplet of two scalar and one pseudoscalar Higgs mesons. In a leading hierarchy approximation, all of the family Higgs are degenerate.  When the texture of the quark masses are taken into account, the degeneracy is lifted into a rich spectrum.  \\

\noindent In this scenario where we have set in the leading hierarchy approximation,  in the {\em diagonal quark mass basis},  $\upsilon_{22}, \,\upsilon_{11} \,=0$, and $\widehat{\upsilon}\,{}_{22}, \,\widehat{\upsilon}\,{}_{11} \,=0$, we have a rich spectrum of Higgs once the texture of quark masses are taken into account.  In this scenario, there are also a family of very high masses. \\

\noindent There is much work that remains to be done to explore the consequences of this proposal.  I presented the preliminary version of this at the conference in honor of Freeman Dyson on the occasion of his 90th birthday in the spirit of exploration and discovery\cite{Chang_Dyson}. \\

\section{Acknowledgments}
\noindent I wish to thank IAS and Nanyang Technological University for the warm hospitality over the consecutive summers where this work was done.  I also wish to thank and acknowledge Zhi-Zhong Xing, Markos Maniatis, and Manmohan Gupta for the many stimulating discussions at IAS on quark texture matrices, and Higgs phenomenology.

\end{document}